\RequirePackage{fix-cm}
\documentclass[smallextended]{svjour3}       
\smartqed  
\usepackage{graphicx}
\begin{document}

\title{Implementation of PhotoZ under Astro-WISE
}
\subtitle{A photometric redshift code for large datasets}


\author{R.P.~Saglia\and
        J.~Snigula\and
        R.~Senger\and
        R.~Bender
}


\institute{Max-Planck Institute for extraterrestrial Physics \at
              Giessenbachstra\ss e, D-85741 Garching, Germany \\
              Tel.: +49-89-300003916\\
              Fax: +49-89-300003495\\
              \email{saglia@mpe.mpg.de}           
           \and
           Universit\"atssternwarte \at
           Scheinerstra\ss e 1, D-81679 M\"unchen, Germany
}

\date{Received: date / Accepted: date}

\maketitle

\begin{abstract}
  We describe the implementation of the PhotoZ code in the framework
  of the Astro-WISE package and as part of the Photometric
  Classification Server of the PanSTARRS pipeline. Both systems allow
  the automatic measurement of photometric redshifts for the millions
  of objects being observed in the PanSTARRS project or expected to be
  observed by future surveys like KIDS, DES or EUCLID.
\end{abstract}

\section{Introduction}
\label{Intro}

Since the completion of the Sloan Digital Sky Survey (SDSS)
\cite{SDSS} optical astronomy has moved on from the detailed studies
of single objects to a phase where catalogues with millions of entries
can be produced. This has allowed for detailed statistical studies of
entire populations, as well as searches for extremely rare objects. 
Consequently, astronomers are forced to update their approach to
data analysis and to embed their codes in database-supported
applications that support efficient automatic procedures and easy
administration of the analysis processes. One such case is the
measurement of photometric redshifts for the hundreds of millions of
galaxies that ongoing or future optical and near infrared
photometric surveys will deliver.

We are directly involved in three large scale imaging surveys.  The
Panoramic Survey Telescope and Rapid Response System 1 (PanSTARRS1,
see \cite{Kaiser04}) project started regular operations in May 2010
and is producing a 5 band (grizy) survey of 3/4 of the sky that at the
end of the forseen 3 years of observations, will be $\approx 1$ mag deeper than
SDSS. Approximately two hundred million galaxies, a similar number of
stars, about a million quasars and $\approx 7000$ Type Ia Supernovae
will be detected. VIKING (VISTA Kilo-Degree Infrared Galaxy Survey
\footnote{http://www.eso.org/public/teles-instr/surveytelescopes/vista/surveys.html})
is a near-infrared 4 band (ZJHK) survey of 1500 square degrees of
extragalactic sky that started in December 2009 at the VISTA
telescope.  This will be complemented in 5 optical bands (ubgri) by
the Kilo Degree Survey (KIDS
\footnote{http://www.strw.leidenuniv.nl/$\tilde{~}$kuijken/KIDS/}) at
the VST telescope, to start in October 2011. Finally, the Dark Energy
Survey (DES \footnote{http://www.darkenergysurvey.org/}) will image
5000 square degrees around the southern galactic pole in 4 optical
bands (bgri) at the CTIO telescope. Looking to the future, we are
participating in the EUCLID
\footnote{http://sci.esa.int/science-e/www/area/index.cfm?fareaid=102}
bid. If approved, the satellite will image 20000 square degrees of the
extragalactic sky in the optical and NIR channels, providing
unprecedented deep photometry for many millions of galaxies.

The science driving these projects ranges from Baryonic Acoustic
Oscillations and growth of structure, to weak shear, galaxy-galaxy
lensing and lensing tomography. All of them rely on the determination
of accurate photometric redshifts for extremely large numbers of
galaxies. Further science goals, like the detection of high redshift
quasars and galaxies, the discovery of very cool stars, or the study of
galaxy evolution with cosmic time will also profit  from the availability
of good photometric redshift and star/galaxy photometric
classification. Therefore, in the last few years we have designed and
implemented schemes to derive and keep organized photometric redshifts,
probability distributions and star/galaxy classification for extremely
large datasets. 

Here we describe two aspects of these efforts; the PhotoZ
implementation for Astro-WISE and the Pan\-STARRS1 Photometric
Classification Server. The structure of the paper is as follows.
Sect. \ref{sec:PhotoZ} considers the algorithm at the core of our
implementations and its recent scientific
use. Sect. \ref{sec:Implementation} discusses the implementation of
the code for large data sets. In Sect. \ref{sec:AWPhotoZ} we present
its Astro-WISE incarnation and give examples of its use and evaluate its
performances in Sect. \ref{sec:AWPerformances}. Sect.  \ref{sec:PCS} is
dedicated to the implementation of the code for the PanSTARRS1
survey. We draw our conclusions in Sect. \ref{sec:Conclusions}.

\section{The PhotoZ code: algorithm and science applications}
\label{sec:PhotoZ}

In the last decade several efficient codes for the determination of
photometric redshifts have been developed and a fair summary of these
efforts would go well beyond the scope of the present contribution.
In short, there are mainly two approaches, one based on empirical
methods, the other on template fitting. In the first case one tries to
parametrize the low-dimensional surface in color-redshift space that
galaxies occupy using low-order polynomials, nearest-neighbor searches
or neural networks \cite{Csabai03,Collister04}. These codes extract the
information directly from the data, given an appropriate training set
with spectroscopic information. Template fitting methods work instead
with a set of model spectra from observed galaxies and stellar
population models \cite{Padmanabhan05,Ilbert06,Mobasher07,Pello09}.

The PhotoZ code that we have implemented under Astro-WISE belongs to
the second category and its original incarnation is described in
\cite{Bender01}. The code estimates redshifts $z$ by comparing a set
of discrete template SEDs T to the broadband photometry of the
(redshifted) galaxies.  For each SED the full redshift likelihood
function including priors for redshift, absolute luminosity and SED
probability is computed using the Bayes' theorem:
\begin{equation}
\label{Bayes}
P(z,T|C,M,...) \propto p(C|z,T)p(z,T|M),
\end{equation}
where $C$ is the vector of measured colors, $M$ the galaxy absolute
magnitude, $p(C|z,T)\propto exp(-\chi^2/2)$ is the probability of
obtaining a normalized $\chi^2$ for the given dataset with its errors,
redshift and template $T$, and $p(z,T|M)$ the prior distribution. This
is a product of parametrized functions of the type:
\begin{equation}
\label{eq:Priors}
p(y)\propto y^nexp\left[-ln(2)\left( \frac{y-\hat{y}}{\sigma_y}\right)^p\right],
\end{equation}
where the variable $y$ stands for redshift or absolute magnitudes.
Typically we use $n=0$, $p=6$ or $8$, and $\hat{y}$ and $\sigma_y$
with appropriate values for mean redshifts and ranges, or mean
absolute magnitudes and ranges, which depend on the SED type. The set
of galaxy templates is semi-empirical and is chosen to map the color
space spanned by the different types of objects at different
redshifts.  The original set \cite{Bender01} includes 31 SEDs
describing a broad range of galaxy spectral types, from early to late
to star-bursting objects. Recently, we added a set of SEDs tailored to
fit luminous red galaxies and one SED to represent the average QSO
spectrum at redshift $\approx 2$ \cite{Saglia11}. Furthermore, the
method also fits a set of stellar templates, allowing a star/galaxy
classification and an estimate of the line-of-sight extinction for
stellar objects.  The templates cover typically the wavelength range
$\lambda=900$ \AA\ up to 25000 \AA\ (with the QSO template covering
instead 300-8000 \AA) and are sampled with a step typically 10 \AA\
wide (varying from 5 to 20 \AA; the QSO SED has $\Delta \lambda=1$
\AA). The method has been extensively tested and applied to several
photometric catalogues with spectroscopic follow-ups.  Given a (deep)
photometric dataset covering the wavelength range from the U to the K
band, excellent photometric redshifts with $\delta z/(1+z)\sim0.03$ up
to $z\approx 5$ with at most a few percent catastrophic failures can
be derived for every SED type (\cite{Gabasch04}, \cite{Feulner05},
\cite{Gabasch08}). When applied to the 5 filter band catalogs $ugriz$
of SDSS \cite{Greisel11} or $grizy$ of PanSTARRS \cite{Saglia11}, the
code delivers $\delta z/(1+z)\sim0.02$ for luminous red galaxies up to
redshift $\approx 0.5$. A more detailed description of the scientific
merits of PhotoZ goes beyond the scope of this paper, see
\cite{Hildebrandt10} to compare these performances to the ones
achieved by other packages. The code is available in Fortran and C++
versions.

\section{Implementation for large datasets}
\label{sec:Implementation}

The science applications described in the previous section dealt with
some thousand objects and could be managed by simple means,
i.e. ascii-based catalogues. In the era of all-sky surveys and/or very
deep fields, where millions, if not billions of objects are imaged,
this approach is doomed to fail. The support of a database, the
automatisation of the procedures and the tools to administrate the
testing and analysis of the results become essential ingredients for a
successful science project. Therefore, having in mind our
participation in the PanSTARRS1 survey and future projects like KIDS,
DES and possibly EUCLID (see Introduction), we
designed and implemented two packages, PhotoZ for Astro-WISE (see
Sect. \ref{sec:AWPhotoZ} and \ref{sec:AWPerformances}) and the
Photometric Classification Server (PCS) for PanSTARRS1 (see
Sect. \ref{sec:PCS}).

\subsection{PhotoZ for Astro-WISE}
\label{sec:AWPhotoZ}

\begin{figure}
\includegraphics[width=12cm]{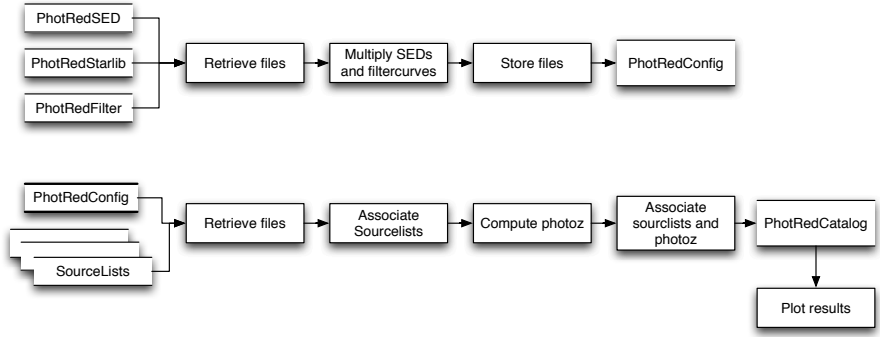}
\caption{Flowchart of the basic functionality of PhotoZ. Top part:
  SEDs, the stellar library and the filter curves are retrieved from
  the data server, the SEDs multiplied with the filter curves to
  compute the relative fluxes in each band. The results are then again
  stored on the data servers. Lower part: To create a PhotRedCatalog
  object, the system retrieves the necessary files, creates an
  AssociateList of the input SourceLists (i.e. matches the lists in RA
  and DEC), computes the photometric redshifts, using the information
  from the PhotRedConfig and finally links the resulting PhotoZ
  SourceList with the AssociateList.}
\label{fig:flowchart}       
\end{figure}

We embedded the PhotoZ code in Astro-WISE following the general
philosophy of the package. A Python wrapper ({\it PhotRedCatalog})
interfaces the Oracle Astro-WISE database to the (Fortran) code,
providing the necessary reading, executing and writing calls to
construct the ascii input files with the photometry vectors, call the
(compiled Fortran) PhotoZ code and transfer the ascii output back into
the database. As usual in every Astro-WISE application, each persistent
entry created in this last phase allows the backward tracing of the
components down to the single raw and calibration frames that went
into the production of the photometry used in the process. The option
for a posteriori evaluation of the full redshift probability
distribution for a list of selected objects is provided. A separate
routine ({\it PhotRedConfig}) allows the folding of the available
spectral energy distributions with the given filter curves on a
predefined grid of redshifts to maximize the speed of the PhotoZ code
for a given photometric set. Parallelization is obtained by splitting
the list of objects to be analysed in smaller chunks, and executing
separate calls of PhotoZ on the multiple cluster nodes. Visualization
routines give the possibility to plot the best-fitting SED, the
best-fitting stellar SED, the datapoints and the redshift probability
distribution of selected objects. A schematic description of the structure of 
the PhotoZ code is given in Fig. \ref{fig:flowchart}

\subsection{Examples and Performances}
\label{sec:AWPerformances}

PhotoZ runs under Astro-WISE as implemented at the Munich node  on the
PanSTARRS cluster, a 175 nodes (each with 2.6GHz 4 CPUs and 6 GB
memory, for a total of 700 CPUs) Beowulf machine with 180 TB disk
space, attached to a PB robotic storing device, mounted at the
Max-Planck Rechenzentrum in Garching. Two servers run the Oracle database. 
The Munich Astro-WISE node is federated with the central node of Groningen.

As an example how the system works, we describe the derivation of the
photometric redshifts of galaxies detected in the Medium Deep Field 4
(MDF04) of PanSTARRS1 (see also \ref{sec:PCS} and \cite{Saglia11}) in
an Astro-WISE session. In this context below we indicate with
``awe$>$'' the Python Astro-WISE prompt. For a detailed description
{\it how to} run the commands discussed below we refer to the
Astro-WISE manual
\footnote{http://www.astro-wise.org/portal/aw\_howtos.shtml}.

We first ingest the PanSTARRS1 filter curves: 
\begin{description}
\item{awe$>$} photredfilter = PhotRedFilter( pathname='PS\_g.filter')
\item{awe$>$} photredfilter.filter=(Filter.mag\_id=='PS\_g g')
\item{awe$>$} ...
\item{awe$>$} photredfilter.make()
\end{description}
where PS\_g.filter is an ASCII file with two columns,
wavelength in Angstroms and the transmission of the PanSTARRS1 g filter at this
wavelength. We repeat the process for the filters r, i, z and y. Then 
we configure the system, specifying the galaxy and stellar libraries (see 
\cite{Bender01}):
\begin{description}
\item{awe$>$} filt = (Filter.name == 'PS\_g')[0]
\item{awe$>$} pfg = (PhotRedFilter.filter == filt )[0]
\item{awe$>$} ...
\item{awe$>$} pse = (PhotRedSED.sed\_name == 'mod\_e.sed')[0]
\item{awe$>$} ps1 = (PhotRedSED.sed\_name == 'mod\_s210.sed')[0]
\item{awe$>$} ...
\item{awe$>$} starlib=(PhotRedStarlib.filename=='starlib\_pickles.lis')[0]
\item{awe$>$} pc = PhotRedConfig()
\item{awe$>$} pc.SEDs=[pse,ps1,...]
\item{awe$>$} pc.filters=[pfg,pfr,pfi,pfz,pfy]
\item{awe$>$} pc.starlib=(starlib)[0]
\item{awe$>$} pc.name='PanSTARRS1\_MDF04'
\item{awe$>$} pc.make()
\end{description}
We now ingest the PanSTARRS1 photometric catalogue into the {\it SourceLists} 
sg, sr, si, sz, sy and generate the photometric redshifts with the commands:
\begin{description}
\item{awe$>$} pr = PhotRedCatalog()
\item{awe$>$} pr.config=pc
\item{awe$>$} pr.master=sg
\item{awe$>$} pr.sourcelists=[sg,sr,si,sz,sy]
\item{awe$>$} pr.name='PanSTARRS1\_MDF04'
\item{awe$>$} pr.make()
\end{description}
The results are stored in the pr.associate\_list AssociateList and can
be examined through the Oracle database tools and/or the Python awe$>$
prompt. For example, the command 
\begin{description}
\item{awe$>$} pr.plot( 23 ) 
\end{description}
plots the best-fitting SED, the best-fitting stellar SED, the
datapoints and the redshift probability distribution for the objects
with identification number 23 in the associate\_list.

The derivation of photometric redshifts for the $\approx 350000$
entries in the MDF04 photometric catalogue down to the $r=24$
magnitude takes 3.3 sec if our full PanSTARRS cluster (700 CPUs) is
available (i.e. $\approx 150$ objects per second per node).  Through
the federation mechanism the results can be seen from each Astro-WISE
federated node that has the relevant permissions to access the
data. We are in the process of optimizing the SEDs and validating the
photometric redshifts through available spectroscopic data for the
PanSTARRS1 filter set. First results are discussed in
\cite{Saglia11}, where a precision of $\approx 0.02(1+z)$ for red
luminous galaxies up to redshift $z\approx 0.5$ is achieved.

\subsection{The Photometric Classification Server for PanSTARRS1}
\label{sec:PCS}

The Photometric Classification Server (PCS) for PanSTARRS1 provides
software tools to perform a photometric star/QSO/galaxy
classification, compute photometric redshifts for galaxies and (a
subset of) best-fitting temperature, metallicity, gravity and
interstellar extinction parameters for stars.  A detailed description
of the system can be found in \cite{Saglia08}, \cite{Snigula09} and
\cite{Saglia11}.  The code is interfaced to the Published Science
Products System (PSPS) database of PanSTARRS1 (see \cite{Heasley08}),
based on Microsoft SQL and inspired in its structure by the SDSS
database.  The ``manual mode'' of operations is similar to the one
described in Sect.  \ref{sec:AWPhotoZ}.  The user can query the
database and run the code off-line through SQL commands and calls to
shell scripts, or also through a web interface. This mode is useful
when optimizing the SEDs using available datasets with spectroscopic
redshifts. The normal mode of operation, however, is fully
automatized. The interface to PSPS triggers the analysis of newly
produced object catalogues of the static sky on a regular basis. When
new entries are found, CAS-like jobs extract them from the database in
Hawaii, format them on our PanSTARRS cluster and submit multiple runs
of the C++ version of PhotoZ with sub-blocks of data to parallelize
the processing. Finally, the resulting photometric redshifts are
stored in a local MySQL database and the corresponding table is pulled
by the central one in Hawaii. The process to download from the PSPS
database the MDF04 catalogue ($\approx 2$ minutes), measure the
photometric redshifts on our PanSTARRS cluster ($\approx 30$ sec if
700 nodes are available) and provide the results for pulling by the
PSPS database ($\approx 2$ minutes) takes at most 5 minutes. Therefore
we expect to sustain the expected regular flow of new photometric data
of PanSTARRS1 without problems.  The system is open for the
implementation of further different approaches to photometric
redshifts (see Sect. \ref{sec:PhotoZ}).

\section{Conclusions}
\label{sec:Conclusions}

PhotoZ under Astro-WISE and PCS for PanSTARRS1, the systems to compute
accurate photometric redshifts for large datasets described in this
paper, are up and running. They are ready to analyse and archive the
photometric catalogues with millions of entries that the wide area
surveys started recently or starting in the near future will provide.
They can be considered as prototypes for the future development of the
data analysis schemes of EUCLID \cite{Laureijs09}.

%
%




\end{document}